\documentclass[prl,10pt,twocolumn]{revtex4-1}
\usepackage{graphicx}
\usepackage{amsmath}
\usepackage{amssymb}
\usepackage{amsthm}
\usepackage{pgfplots}

\begin{document}

\title{Generating Robust Optical Entanglement in Weak Coupling
Optomechanical Systems}

\author{Mark C. Kuzyk}
\email{mkuzyk@uoregon.edu}
\author {Steven J. van Enk}
\author {Hailin Wang}
\affiliation{Department of Physics and Oregon Center for Optics, University
of Oregon, Eugene, OR 97403, USA}
\date{\today}

\begin{abstract}
A pulsed scheme for generating robust optical entanglement via the
coupling of two optical modes to a mechanical oscillator is proposed.
This scheme is inspired by the S{\o}rensen-M{\o}lmer approach for
entangling trapped ions in a thermal environment and is based on the use
of optical driving pulses that are slightly detuned from the respective
sideband resonance.  We show that for certain pulse durations, the
optomechanical interaction can return the mechanical oscillator to its
initial state.  The corresponding entanglement generation is robust
against thermal mechanical noise in the weak as well as the strong
coupling regimes.  Significant optical entanglement can be generated in
the weak coupling regime, even in the presence of a large thermal phonon
occupation.
\end{abstract}
\maketitle


\textit{Introduction-}  In an optomechanical resonator, a mechanical
oscillator can interact with any of the optical modes via radiation
pressure.  This property can enable a quantum interface that converts
photons between vastly different wavelengths or couple together
different types of quantum systems in a hybrid quantum network
\cite{Stannigel2010,Tian2010,Safavi2011.1,Regal2011,Wang2012,Tian2012,Singh2012,Dong2012,Hill2012,Liu2013,Mcgee2013}.
A multi-mode optomechanical system also provides an experimental
platform for generating continuous variable quantum entanglement of
optical modes through the formation of Bogoliubov optical modes, and in
particular, for generating entanglement between optical and microwave
modes \cite{Barzanjeh2012,Tian2013,Wang2013,Barzanjeh2011}.

Entanglement generation is often hampered by dissipation and decoherence
induced by the unavoidable coupling to the environment.  For generation
of optical entanglement via an optomechanical process, a major obstacle
is the coupling of the mechanical oscillator to the thermal reservoir.
A recently proposed scheme has exploited the coherent dynamics of the
Bogoliubov modes to circumvent thermal mechanical noise \cite{Tian2013}.
The thermal robustness of the Bogoliubov-mode based schemes hinges on
the achievement of very strong optomechanical coupling, for which the
effective multi-photon optomechanical coupling rate far exceeds the
damping rates of the relevant optical and mechanical modes.  Although
strong coupling has been achieved for individual optomechanical systems
in both optical and microwave regimes
\cite{Groblacher2009,Teufel2011,Verhagen2012}, it is
exceedingly difficult to have the optomechanical coupling rate to be much greater than the cavity decay rate in the
optical regime, especially in a setting that is suitable for generating
entanglement between optical and microwave modes.

In this letter, we propose and analyze an optomechanical scheme for
optical entanglement generation, which takes advantage of a special
class of multi-mode interaction Hamiltonian, instead of Bogoliubov
modes, to circumvent thermal mechanical noise.  This scheme is inspired
by earlier theoretical and experimental studies on entangling trapped
ions in a thermal environment
\cite{sorensen1999,Sackett2000,Sorensen2000}.  In these studies, the
entanglement operation takes place via the mechanical degrees of freedom
of the ions.  As shown by S{\o}rensen and M{\o}lmer, robust entanglement
can be achieved in a thermal environment with a class of Hamiltonian
that returns the motion of the ions to their initial state upon the
completion of the entanglement operation
\cite{Sackett2000,Sorensen2000}.  Here, we outline a pulsed entanglement
scheme using an optomechanical interaction Hamiltonian that has the
features of the S{\o}rensen-M{\o}lmer (S-M) mechanism.  The entanglement
scheme, which will be referred to as the S{\o}rensen-M{\o}lmer scheme,
can function in the weak as well as strong coupling regime. In
comparison with the Bogoliubov-mode based schemes, the
S{\o}rensen-M{\o}lmer scheme can remain robust against the thermal
mechanical noise even in the weak coupling regime.  Our theoretical
analysis shows that significant optical entanglement can be generated in
the weak coupling regime, even in the presence of a large thermal phonon
occupation ($n_{th} \sim 1000$).


\textit{Three-mode optomechanical system-}  We consider an
optomechanical system, in which two optical modes with resonance
frequencies $\omega_{c,i}$ ($i=1,2$) and cavity linewidths $\kappa_i$,
couple to a mechanical oscillator of frequency $\omega_m$ and mechanical
linewidth $\gamma$ (see Fig. 1a). The optomechanical coupling is driven
by strong laser fields of frequency $\omega_{L,i}$ near the mechanical
sideband of the respective cavity resonance.  This type of three-mode
optomechanical systems has already been used for the experimental
demonstration of optomechanics-based optical wavelength conversion
\cite{Dong2012,Hill2012,Liu2013} and for the realization of an
optomechanical dark mode \cite{Dong2012, Massel2012}.  In a frame where
each optical mode rotates at its driving frequency $\omega_{L,i}$, and
after the standard linearization process, the effective Hamiltonian of
the system is
\begin{equation}\label{linham}
H = \omega_{m}b^{\dagger}b + \sum_{i=1}^{2}
\left( \Delta_{i}a_{i}^{\dagger}a +
g_{i}(a_{i}+a_{i}^{\dagger})(b+b^{\dagger})
\right),
\end{equation}
where $b$ and $a_i$ are the annihilation operators for the mechanical
and optical modes, respectively, and $\Delta_i = \omega_{c,i} -
\omega_{L,i}$ is the detuning of the driving field from the respective
cavity resonance.  The effective coupling rate $g_i$ is controlled by
the strength of the driving field according to $g_i =
\sqrt{N_i}g_{0,i}$, where $N_i$ is the intra-cavity photon number for
the driving field and $g_{0,i}$ is the single-photon optomechanical
coupling rate.

The linearized interaction Hamiltonian couples each optical mode to the
mechanical oscillator with two types of interaction.  A beam-splitter
interaction, associated with the term $g_i(a_{i}^{\dagger}b +
a_{i}b^{\dagger})$, is an anti-Stokes scattering process that can enable
state transfer between optical and the mechanical systems.  A two-mode
squeezing interaction, of the form
$g_i(a_{i}b+a_{i}^{\dagger}b^{\dagger})$, is a Stokes scattering process
that generates correlated phonon-photon pairs. The beam-splitter
interaction has been used for the experimental realization of coherent
inter-conversion between optical and mechanical excitations
\cite{Verhagen2012,Fiore2013,Palomaki2013} as well as the
optomechanically-induced transparency
\cite{Teufel2011,Weis2010,Safavi2011.2} and has also been exploited for
optical wavelength conversion in the three-mode optomechanical system.
The two-mode squeezing interaction has been employed in earlier
theoretical proposals for generating continuous variable entanglement
between optical and mechanical modes and also between two mechanical
modes \cite{Mancini2002,Paternostro2007,Hofer2011,Tan2013, Woolley2013,
Furusawa1998}.

For the generation of two-mode optical entanglement, mode 1 is driven
near the red sideband, at frequency $\omega_{L,1} = \omega_{c,1} -
\omega_m - \Delta$, while mode 2 is driven near the blue side-band, at
frequency $\omega_{L,2} = \omega_{c,2} +\omega_m + \Delta$, where
$\Delta$ is the detuning from the sideband resonance, as illustrated
schematically in Fig. 1b. The optomechanical system is assumed to be in
the resolved sideband limit, with $\omega_{m} \gg \kappa_{1,2} \gg
\gamma$, such that a driving field near the red sideband or blue
sideband drives either the beam-splitter or two-mode squeezing
interaction, respectively. Heuristically, entanglement between modes 1
and 2 in this system is generated in two steps.  The two-mode squeezing
interaction driven by the laser field near the blue sideband generates
entanglement between phonons in the mechanical oscillator and photons in
mode 2.  The beam-splitter interaction driven by the laser field near
the red sideband then maps the state of the entangled phonons onto
photons in mode 1.


\textit{S{\o}rensen-M{\o}lmer Mechanism-}  To gain insights into the
dynamics of the coherent optomechanical interactions and to discuss the
S-M mechanism for the three-mode optomechanical system,
we first ignore the damping of both optical and mechanical systems and
adjust the optomechanical coupling rates for the two optical modes such
that $g_{1} = g_{2} = g$.  In this limit, the interaction Hamiltonian for
the entanglement generation falls into a class discussed originally by
M{\o}lmer and S{\o}rensen and also by Milburn
\cite{sorensen1999,Sorensen2000, Milburn1999}.  For this class, the
exact propagator can be written in a form
\begin{equation}
U(t) = e^{-iA(x,p,t)}e^{-iF(x,p,t)x_b}e^{-iG(x,p,t)p_b},
\end{equation}
where $x = x_1 + x_2$ and $p = p_2 - p_1$ are EPR-like variables, with
the dimensionless quadrature variables defined as $x_i = (a_i +
a_{i}^{\dagger})/\sqrt{2}$, $p_i = i(a_{i}^{\dagger} - a_i )/\sqrt{2}$,
and similarly for the mechanical mode operators $x_b$ and $p_b$.  At
regularly spaced time intervals $t_n = 2\pi n/\Delta$,
\begin{equation}
F(x,p,t_n) = G(x,p,t_n) = 0,
\end{equation}
returning the mechanical degrees of freedom to their initial states.  At
the same time, $A(x,p,t_n)$, which is given by,
\begin{equation}
A(x,p,t_n) = -\frac{g^2}{2\Delta}(x^2+p^2)t_n
\end{equation}
generates entanglement between modes 1 and 2, according to
\begin{equation}
U^{\dagger}(x,p,t_n)a_{1(2)}U(x,p,t_n) = \mu a_{1(2)} + \nu
a_{2(1)}^{\dagger},
\end{equation}
where $\mu = 1+ir$ and $\nu = ir$, with a squeezing parameter $r =
g^{2}t_{n}/2\Delta$ (see the supplementary materials for the derivation
of the propagator and for the analytical expression of the
entanglement).

It is remarkable that independent of the particular form of the initial
state of the system, the mechanical oscillator periodically returns to
its initial state, and leaves the optical modes increasingly entangled
upon each return.  The entanglement is generated through the mechanical
motion of the system. However, the final entangled optical state
contains no information of the mechanical system, and can thus be robust
against thermal Brownian noise that enters the system through the
mechanical oscillator.  Note that in the limit of large detuning
$\Delta$, the mechanical degrees of freedom can be adiabatically
eliminated.  The optical entanglement generation can thus become
thermally robust without satisfying the condition, $t_n = 2\pi
n/\Delta$.  The large detuning, however, limits the amplitude of the
squeezing parameter and hence the degree of entanglement that can be
achieved.

\textit{Analysis with Langevin equations-}  We have used the quantum
Langevin equations to analyze in detail the dynamics of the entanglement
generation and especially the effects of thermal mechanical noise.  We
work in a rotating frame $\tilde{H}=U_{R}HU_{R}^{\dagger}$, where $H$ is
the Hamiltonian of Eq. (\ref{linham}), and $U_{R}=e^{i(\omega_m +
\Delta)(a_{1}^{\dagger}a_{1} - a_{2}^{\dagger}a_{2} + b^{\dagger}b)t}$.
In this frame, the quantum Langevin equations in the resolved sideband
limit have the form
\begin{align}\label{langevin}
\dot{a}_{1} &= -\frac{\kappa_1}{2}a_1 - ig_1b - \sqrt{\kappa_1}a_{in,1} \\
\dot{a}_{2}^{\dagger} &= -\frac{\kappa_2}{2}a_{2}^{\dagger} + ig_2b - \sqrt{\kappa_2}a_{in,2}^{\dagger} \\
\dot{b} &= -(i\Delta + \frac{\gamma}{2})b -ig_1a_1 - ig_2a_{2}^{\dagger}
- \sqrt{\gamma}b_{in},
\end{align}
where the resolved sideband limit has allowed us to drop all
counter-rotating terms.  The input operators for the optical modes,
$a_{in,i}(t)$, characterize the optical cavity coupling to the vacuum,
and have correlation functions $\langle
a_{in,i}(t)a_{in,i}^{\dagger}(t^{\prime}) \rangle =
\delta(t-t^{\prime})$.  The Brownian noise that enters the system
through the mechanical degree of freedom is described by the operator
$b_{in}(t)$.  We assume the system to have a sufficiently large
mechanical quality factor $Q_m = \omega_m/\gamma$ such that the Brownian
noise can be approximated to be Markovian \cite{Genes2008}, with $\langle
b_{in}(t)b_{in}^{\dagger}(t^{\prime}) \rangle =
(n_{th}+1)\delta(t-t^{\prime})$.

The entanglement is generated for optical driving pulses with a given
duration and is quantified with the logarithmic negativity,
$E_\mathcal{N}$ \cite{Plenio2005, Vidal2001}.  We limit the duration of
the optical pulse to ensure that the optomechanical system remains
dynamically stable and that nonlinear optomechanical interactions are
negligible.  For typical optomechanical systems, the mechanical damping
rate can be much smaller than both the cavity linewidth and the
effective optomechanical coupling rate. To generate strong entanglement
and maintain thermal robustness, we have used sideband detuning that is
less than $g$, but far exceeds $\gamma$. In the following, we first
consider the intracavity entanglement in the
strong coupling regime, where $g \gg \kappa_i$.  We then analyze the
entanglement contained in an output mode for a system in the bad cavity
limit with $g \ll \kappa_i$.     

\textit{Strong coupling-}  Figure 2a plots the intracavity entanglement
generated in the strong coupling regime.  As shown in Fig. 2a,
the negativity oscillates as a function of time, with the peaks or the
maxima of the negativity located at times $t_n$, when the mechanical
degree of freedom is returned to its initial state, as anticipated from
the above theoretical treatment without the inclusion of the damping
processes.  With increasing thermal phonon occupation, the maxima
decrease gradually, but the oscillation becomes much more pronounced,
with the minima in the negativity quickly approaching zero, illustrating
the importance and also the effectiveness of the S-M mechanism in
circumventing the thermal mechanical noise.

For comparison, Fig. 2b plots the intracavity entanglement as a function
of time, generated in the same system and under otherwise similar
conditions by the method of the Bogoliubov mode \cite{Tian2013}.  In
this case, the entanglement maxima occur when the mechanical oscillator
returns to its initial state through the Rabi oscillation of the bright
Bogoliubov modes that couple to the mechanical oscillator. The period of
the oscillation in the negativity in Fig. 2b is thus determined by the
effective optomechanical coupling rate of the bright modes.  At very low
thermal phonon occupation, the Bogoliubov mode approach can generate
stronger maximum entanglement.  However, the entanglement is much more
sensitive to the timing of the optical field than that generated with
the S-M mechanism (see Fig. 2). A small deviation from an exact
optomechanical $\pi$ pulse leads to appreciable mixing between the
optical and mechanical excitations.

For a more detailed comparison of the thermal robustness of the two
entanglement schemes, we plot in Fig. 3 the maximum negativity obtained
under the conditions of Fig. 2 for each entanglement scheme as a
function of the initial thermal phonon occupation.  As shown in Fig. 3,
the S{\o}rensen-M{\o}lmer scheme becomes advantageous when $n_{th}$
exceeds 500, which further highlights the robustness of the
S{\o}rensen-M{\o}lmer scheme against thermal mechanical noise.  Because
of the detuning from the sideband resonance, the S-M mechanism is more
effective in returning the mechanical oscillator to its initial state
than the coherent dynamics of the Bogoliubov bright modes and thus is
more robust against thermal mechanical noise.

\textit{Bad cavity limit-} In the bad cavity limit, we
solve the optical modes adiabatically and investigate the entanglement
in the output of the cavity as a function of pulse duration. The
entanglement in the cavity output is more relevant to experimental
implementation and to potential applications than the intracavity
entanglement. Starting with Eq. (\ref{langevin}), the adiabatic
solutions for the optical modes are
\begin{align}
a_1(t) &= -\frac{2ig_1}{\kappa_1}b(t) -
\frac{2}{\sqrt{\kappa_1}}a_{in,1}(t) \\
a_{2}^{\dagger}(t) &= \frac{2ig_2}{\kappa_2}b(t) -
\frac{2}{\kappa_2}a_{in,2}^{\dagger}(t),
\end{align}
where $b(t)$ is the formal solution of the mechanical mode.  Using the
input-output relation $a_{out}=a_{in}+\sqrt{\kappa}a$, the cavity
output is related to the input by
\begin{align}
a_{out,1}(t) &= -2i\sqrt{G_1}b(t) - a_{in,1}(t) \\
a_{out,2}^{\dagger}(t) &= 2i\sqrt{G_2}b(t) - a_{in,2}^{\dagger}(t),
\end{align}
where
\begin{widetext}
\begin{equation}
b(t) =
b(0)e^{-zt}+e^{-zt}\int_{0}^{t}e^{zs}\left(
2i\sqrt{G_1}a_{in,1}(s)
+2i\sqrt{G_2}a_{in,2}^{\dagger}(s)
- \sqrt{\gamma}b_{in}(s) \right)\mathrm{d}s.
\end{equation}
\end{widetext}
The complex number $z = \Gamma + i\Delta$ contains an effective damping
rate $\Gamma = 2G_1 - 2G_2 + \gamma/2$, where the coupling rates $G_i =
g_{i}^2/\kappa_i$ effectively characterize the optomechanical
interaction strength in the bad cavity limit.  This also leads to a
modified requirement for the S-M mechanism, $G_1 = G_2$.

The output modes $a_{out,i}(t)$ are improper continuous operators, not
well suited for characterizing entanglement.  One may instead describe
the system in a discrete mode basis by defining independent discrete
bosonic operators \cite{Blow1990}
\begin{equation}\label{discrete}
A_{out,i}^{(k)} = \int\mathrm{d}t \hspace{1mm} \phi_{k}^{*}(t)a_{out,i}(t)
\end{equation}
where $i=1,2$ again label the two optical modes of the system, the index $k$
labels members of a denumerably infinite set, and the mode functions
$\phi_k(t)$ form a complete orthonormal basis under the inner product
$\int \mathrm{d}t \hspace{1mm} \phi_{k}^{*}(t)\phi_{k^{\prime}}(t)$.
The operators defined by equation (\ref{discrete}) satisfy the proper
commutation relations, $[A_{out,i}^{(j)},A_{out,i}^{(k)\dagger}] =
\delta_{jk}$, for characterizing the entanglement of the output modes with
logarithmic negativity.

We study the entanglement between two particular discrete modes of the
output, defined as
\begin{equation}\label{discrete_modes}
A_{out,i} = \frac{1}{\sqrt{\tau}}\int_{0}^{\tau}\mathrm{d}t \hspace{1mm}
a_{out,i}(t).
\end{equation}
These modes have central frequencies at the cavity resonances
$\omega_{c,1}$ and $\omega_{c,2}$, and describe pulses of duration
$\tau$.  By extracting only one mode from each output field, we have
performed a local operation, which can only decrease the total amount of
entanglement in the system \cite{Vidal2000,van2003}.  Thus, the
entanglement we calculate gives a lower bound on the total entanglement
of the system.

The S-M mechanism remains effective in the regime of weak optomechanical
coupling.  Figure 4a plot the entanglement contained in the modes
defined by Eq. (\ref{discrete_modes}), as a function of the pulse duration
$\tau$, and for various thermal phonon occupations.  Similar to the
results obtained in the strong coupling regime shown in Fig. 2a, we find
that the negativity oscillates with the pulse duration, with the
entanglement maxima occurring at pulse durations satisfying the
condition of $t_n = 2\pi n/\Delta$.  With increasing thermal phonon
occupation, the maxima decrease gradually, while the minima quickly
approach zero.  Significant entanglement can be still achieved with a
thermal phonon occupation of order 1000.

The S-M mechanism for the three-mode optomechanical system requires
equal effective optomechanical coupling for the two optical modes. To
illustrate this, we plot in Fig. 4b the negativity as a function of the
thermal phonon occupation when the requirement of $G_1 = G_2$ is
satisfied (solid), and when the requirement is not (dashed).  Thermally
robust entanglement can be achieved only when $G_1 = G_2$ is
satisfied.  Note that with $\Delta = 0$ and in the weak
coupling regime, thermally robust entanglement cannot be achieved
regardless whether $G_1 = G_2$ is satisfied.

\textit{Conclusions-}  In summary, we have presented a pulsed approach,
in which the optical driving fields are slightly detuned from the
respective sideband resonance, for generating optical entanglement in a
three-mode optomechanical system.  In this approach, the mechanical
oscillator returns to its initial state and is disentangled with the
optical modes upon the completion of the entanglement operation.
Although schemes based on the use of the Bogoliubov modes
can lead to greater entanglement when the mechanical oscillator is near
the motional ground state, the S{\o}rensen-M{\o}lmer scheme is more robust
against thermal mechanical noise.  In particular, significant
entanglement can still persist at relatively high thermal phonon
occupation in the weak coupling regime, providing a promising avenue for
generating optical entanglement, including that between optical and microwave modes.

We would like to thank Lin Tian for helpful discussions and valuable assistances.  This work is supported by NSF award No. 1205544 and by the DARPA-MTO ORCHID program through a grant from AFOSR.

\begin{figure}
\includegraphics[width=0.8\linewidth,keepaspectratio]{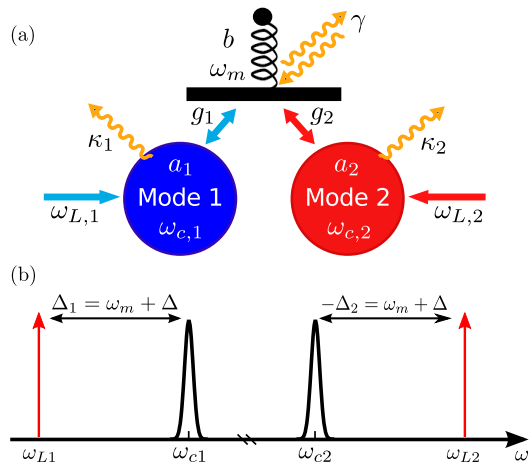}
\caption{(a) Schematic of the three-mode optomechanical system. (b)
Spectral position of the optical driving fields.}
\label{fig1}
\end{figure}

\begin{figure}
\begin{minipage}[t]{0.23\textwidth}
\includegraphics[width=\linewidth,height=31mm]{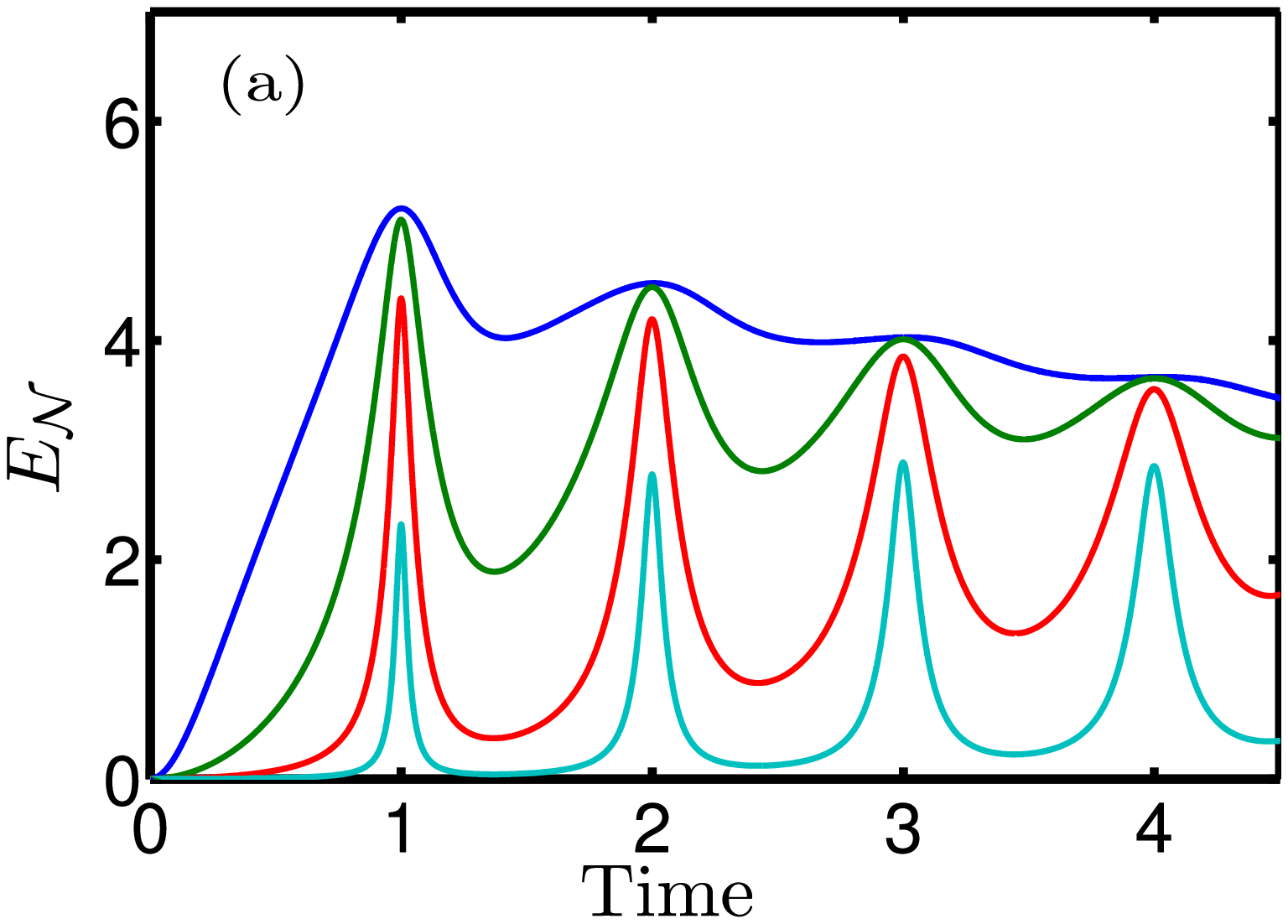}
\end{minipage}
\begin{minipage}[t]{0.23\textwidth}
\includegraphics[width=\linewidth,height=32.5mm]{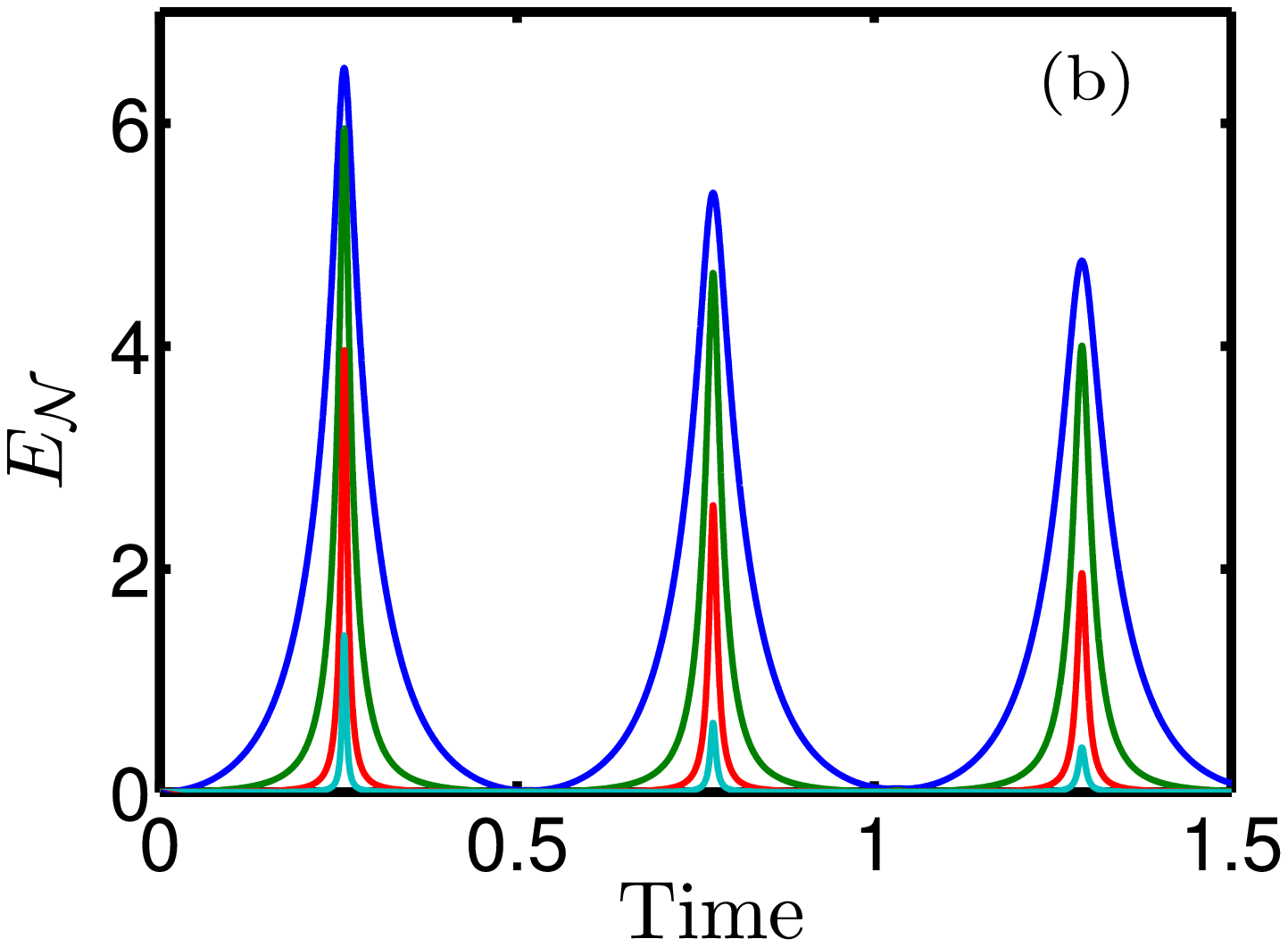}
\end{minipage}
\caption{Intracavity entanglement versus time.  (a)
S{\o}rensen-M{\o}lmer scheme with $g/\gamma = 4\cdot10^3$ and
$\Delta/\gamma = 10^3$.  (b) Bogoliubov mode scheme with $g_1/\gamma =
4\cdot10^3$ and $g_2/\gamma = 3.5\cdot10^3$.  For both (a) and (b),
$\kappa_1/\gamma = \kappa_2/\gamma = 10$ and the time is in units of
$2\pi/(10^3\gamma)$. From top to bottom, $n_{th} = 10, 10^2, 10^3,
10^4$.}
\label{fig2}
\end{figure}

\begin{figure}
\includegraphics[height=50mm]{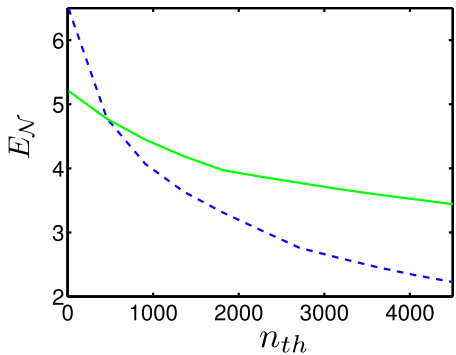}
\caption{Maximum intracavity entanglement as a function of thermal
phonon occupation $n_{th}$.  The solid (dashed) line is for the
S{\o}rensen-M{\o}lmer (Bogoliubov mode) scheme.}
\label{strong_nth}
\end{figure}

\begin{figure}
\begin{minipage}[t]{0.23\textwidth}
\includegraphics[width=\linewidth]{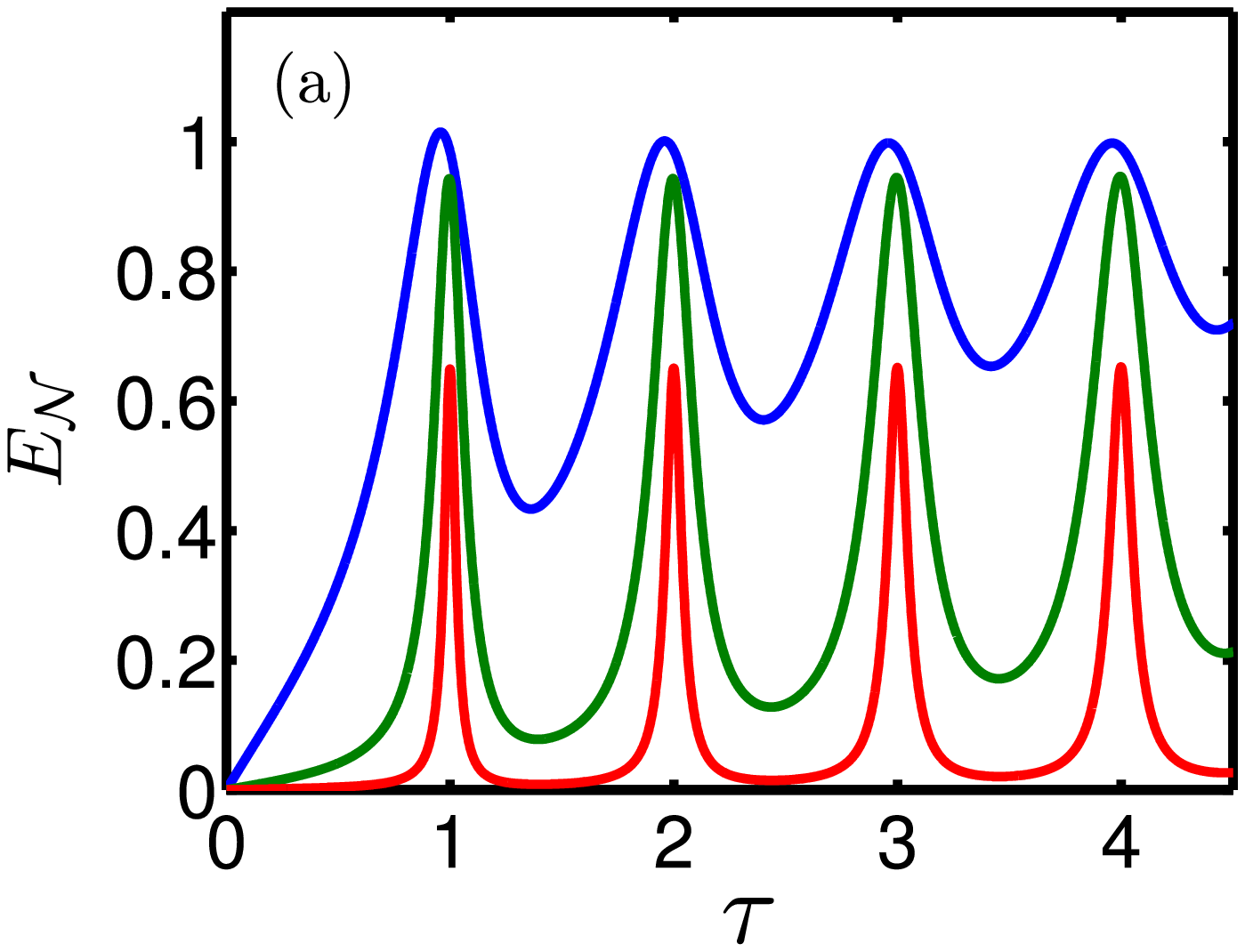}
\end{minipage}
\begin{minipage}[t]{0.23\textwidth}
\includegraphics[width=\linewidth,height=31.5mm]{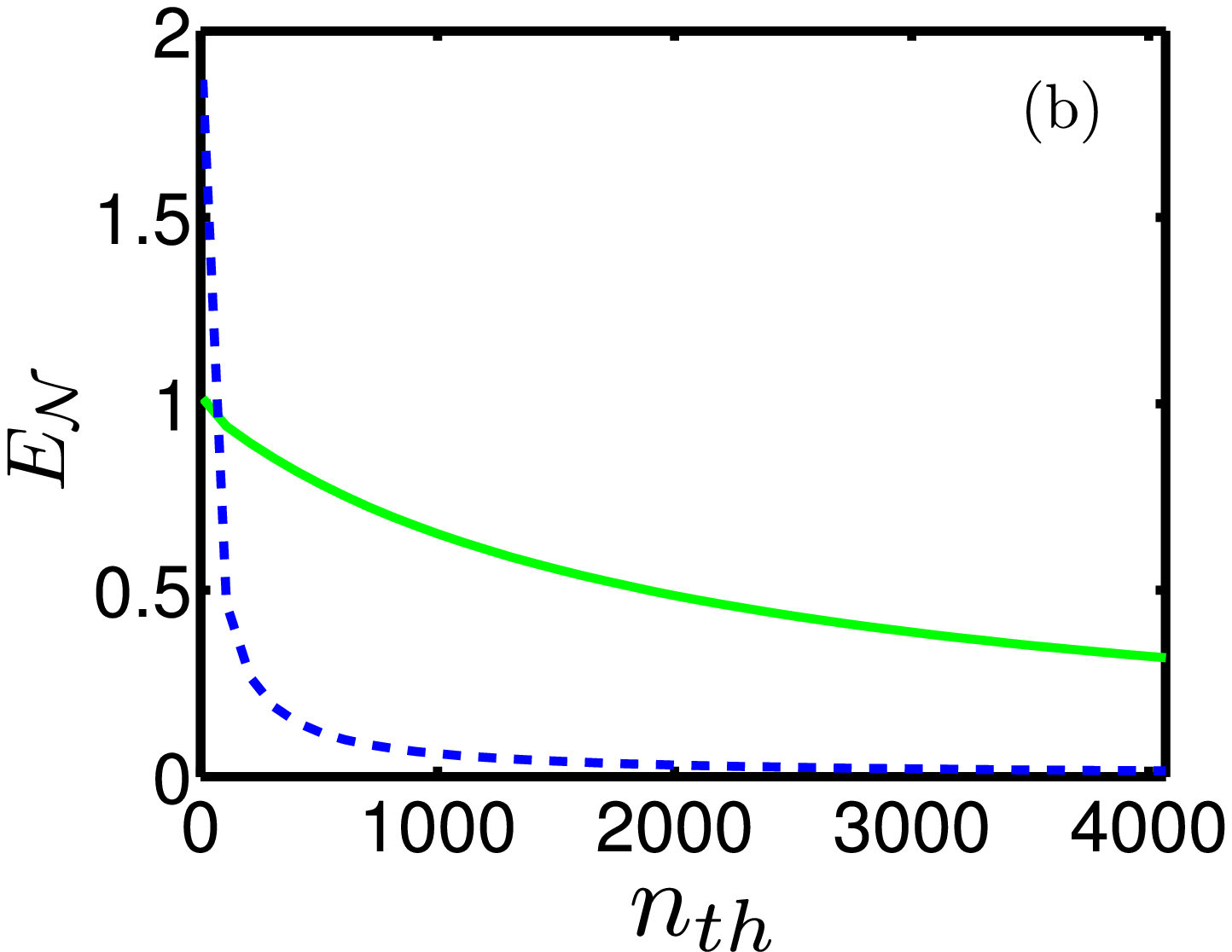}
\end{minipage}
\caption{Entanglement of an output mode in the bad cavity limit, with
$\Delta/\gamma = 10^3$ and $\kappa_1/\gamma = \kappa_2/\gamma = 6\cdot10^3$.  (a) As a function of pulse duration $\tau$, in
units of $2\pi/(10^3\gamma)$, with $G_1/\gamma = G_2/\gamma = 667$.
From top to bottom, $n_{th}=10, 10^2,10^3$.  (b) Maximum entanglement generated as a function of thermal phonon occupation. Solid line: $G_1 =
G_2$. Dashed line: $G_1/\gamma = 667$ and $G_2/\gamma = 540$.}
\label{fig3}.
\end{figure}

\end{document}


\title{Supplementary Materials}

\author{Mark C. Kuzyk}
\email{mkuzyk@uoregon.edu}
\author {Steven J. van Enk}
\author {Hailin Wang}
\affiliation{Department of Physics and Oregon Center for Optics, University
of Oregon, Eugene, OR 97403, USA}

\maketitle

\section{Unitary Evolution}

In this section we discuss the evolution of the three-mode
optomechanical system, neglecting all damping terms.  The interaction
Hamiltonian for the system is
\begin{equation}
H_I = (g_1 a_1 + g_2a_{2}^{\dagger})b^{\dagger}e^{i\Delta t}
+ \mathrm{H.c.}
\end{equation}
We assume from now on that the optomechanical coupling rates for the two
optical modes are set equal, $g_1 = g_2 \equiv g$.  We define
dimensionless quadrature variables $x_i = (a_i +
a_{i}^{\dagger})/\sqrt{2}$, $x_b = (b + b^{\dagger})/\sqrt{2}$, $p_i =
i( a_{i}^{\dagger} - a_i)/\sqrt{2}$, and $p_b =
i(b^{\dagger}-b)/\sqrt{2}$.  From the optical field quadratures, we
define two EPR variables $x \equiv x_1 + x_2$, and $p \equiv p_2 - p_1$,
which satisfy $[x,p]=0$ and can therefore be treated as numbers for the
current treatment.  In terms of these variables, the interaction
Hamiltonian can be written in the form 
\begin{equation}
H_{I} = f(t)x_b + g(t)p_b.
\end{equation}
The time-dependent coeffecients of the mechanical degrees of freedom are 
\begin{align}
f(t) &= g[x\cos(\Delta t) + p\sin(\Delta t)] \\
g(t) &= g[x\sin(\Delta t) - p \cos(\Delta t)].
\end{align}

We write the exact propagator by ansatz, assuming the form
\begin{equation}
U(t) = e^{-iA(t)}e^{-iF(t)x_b}e^{-iG(t)p_b},
\end{equation}
and solve for the functions $A(t), F(t),$ and $G(t)$ by enforcing that
$U(t)$ satisfy the equation
\begin{equation}
i\frac{d}{dt}U(t) = H_{I}U(t).
\end{equation}
In doing so, one finds the the solutions
\begin{align}
F(t) &= \int_{0}^{t}\mathrm{d}t^{\prime}f(t^{\prime}) \nonumber \\
G(t) &= \int_{0}^{t}\mathrm{d}t^{\prime}g(t^{\prime})  \\
A(t) &= -\int_{0}^{t}\mathrm{d}t^{\prime}F(t^{\prime})g(t^{\prime}) \nonumber 
\end{align}
Following through the integration yields
\begin{align}
F(t) &= \frac{g}{\Delta}[x\sin(\Delta t) - p \cos(\Delta t) + p]
\nonumber \\
G(t) &= \frac{g}{\Delta}[x - x\cos(\Delta t) - p \sin(\Delta t)]
\end{align}
and
\begin{align}
A(t) = &-\frac{g^2}{\Delta^2}\Big( \frac{t\Delta}{2}(x^2 + p^2) 
\nonumber \\
&+ \frac{1}{4}\sin(2\Delta t)(p^2 - x^2) 
+ \frac{px}{2}[\cos(2\Delta t) -
1] \nonumber \\ 
&- px[\cos(\Delta t) - 1] - p^2 \sin(\Delta t)\Big).
\end{align}
The coefficients of the mechanical degrees of freedom oscillate in time,
simultaneously returning to zero whenever the timing condition $t_n =
2\pi n/\Delta$ for integer $n$ is satisfied.  At those times, the
remaining part of the propagator entangles the optical modes with the
operation of
\begin{equation}
A(t_n) = -\frac{g^2}{\Delta^2}\pi n (x^2 + p^2).
\end{equation}
For optical states initially in the vacuum, the covariance matrix of the
optical modes can be constructed, and a detailed calculation gives the
logarithmic negativity
\begin{equation}
E_{\mathcal{N}} = -\frac{1}{2}\log_2\left( 2r^2 - \sqrt{4r^8 + 8r^6 +
5r^4 + r^2} + 2r^4 + \frac{1}{4}  \right) - 1
\end{equation}
where $r = \pi n g^2 / \Delta^2$.

\section{Logarithmic Negativity}

To quantify the entanglement between the optical modes of the system, we
use the logarithmic negativity.  For two-mode Gaussian states described
by annihilation operators $a_{i}$ ($i=1,2$) that satisfy the bosonic
commutation relations $[a_i,a_{j}^{\dagger}] = \delta_{ij}$, the
logarithmic negativity can be calculated from the expression
\begin{equation}\label{logneg}
E_{\mathcal{N}} = \mathrm{max}\left( 0, -\log_2 2\eta^{-}  \right),
\end{equation}
where 
\begin{equation}
\eta^{-} = \frac{1}{\sqrt{2}}\sqrt{\Sigma - \sqrt{\Sigma^2 -
4\mathrm{det}V}},
\end{equation}
and
\begin{equation}
\Sigma = \mathrm{det}A + \mathrm{det}B - 2\mathrm{det}C.
\end{equation}
The matrices $A, B,$ and $C$ are $2\times 2$ blocks of the
covariance matrix
\begin{equation}
V = \left(
\begin{array}{cc}
A & C \vspace{1mm} \\
C^{\mathrm{T}} & B
\end{array}
\right).
\end{equation}
The components of the covariance matrix have the usual form
\begin{equation}
V_{ij} = \frac{1}{2}\langle \Delta \xi_i \Delta \xi_j + \Delta \xi_j
\Delta \xi_i \rangle,
\end{equation}
where $\Delta \xi_i = \xi_i - \langle \xi_i  \rangle$, and $\vec{\xi} =
[x_1, p_1, x_2, p_2]^{\mathrm{T}}$.  The dimensionless quadrature variables
$x_i$ and $p_i$ are constructed from the annihilation operators
according to $x_i = (a_i + a_{i}^{\dagger})/\sqrt{2}$ and $p_{i} =
i(a_{i}^{\dagger} - a_i)/\sqrt{2}$.

From Eq. (\ref{logneg}), one finds that the system becomes entangled when 
$\eta^{-}<1/2$.  In terms of the covariance matrix, the requirement for
entanglement is $4\det V < \Sigma - 1/4$, which is equivalent to Simon's
partial transpose criterion \cite{Simon2000}. 

\bibliography{biblio}